%% file: DGrupe.tex
\documentclass[preprint2]{aastex}

\input{psfig}
\usepackage{epsf}
\usepackage{graphics}

\newcommand{\kms}{km s$^{-1}$}

\newcommand{\rb}[1]{\raisebox{1.5ex}[-1.5ex]{#1}}
\newcommand{\msun}{$M_{\odot}$}

\newcommand{\pl}{$\pm$}

\newcommand{\sig}{$\sigma$}
\newcommand{\mbh}{$M_{\rm BH}$}
\newcommand{\etal}{{\it et al.}~}
\shorttitle{\mbh--$\sigma$ relation}
\shortauthors{Grupe \& Mathur}

\begin{document}

\input DGrupe_clipfig.tex
\useunitmm

\def \charthoffset {\hspace{0.2cm}} \def \charthsep {\hspace{0.3cm}}
\def \chartvsepcap {\vspace{0.3cm}}
\def \chartvsep {\vspace{0.1cm}}
\newcommand{\putchartb}[1]{\clipfig{/home/halley/dgrupe/ps/#1}{85}{20}{0}{275}{192}}
\newcommand{\putchartc}[1]{\clipfig{/home/halley/dgrupe/ps/#1}{55}{33}{19}{275}{195}}
\newcommand{\chartlineb}[2]{\parbox[t]{18cm}{\noindent\charthoffset\putchartb{#1}\charthsep\putchartb{#2}\chartvsep}}

\newcommand{\chartlinec}[2]{\parbox[t]{18cm}{\noindent\charthoffset\putchartc{#1}\charthsep\putchartc{#2}\chartvsep}}


\title{\mbh--$\sigma$ relation for a Complete Sample of Soft X-ray Selected 
AGNs
}


\author{Dirk Grupe 
and Smita Mathur}
\affil{Astronomy Department, Ohio State University,
    140 W. 18th Ave., Columbus, OH-43210, U.S.A.}
\email{dgrupe@astronomy.ohio-state.edu}




\begin{abstract}

We present black hole mass--bulge velocity dispersion relation for a
complete sample of 75 soft X-ray selected AGNs: 43 broad line Seyfert
1s and 32 narrow line Seyfert 1s. We use luminosity and FWHM(H$\beta$)
as surrogates for black hole mass and FWHM([OIII]) as a surrogate for
the bulge velocity dispersion. We find that NLS1s lie below the
\mbh--\sig\ relation of BLS1s, confirming the Mathur \etal (2001)
result. The statistical result is robust and not due to any systematic
measurement error. This has important consequences towards our
understanding of black hole formation and growth: black holes grow by
accretion in well formed bulges, possibly after a major merger. As
they grow, they get closer to the \mbh--\sig\ relation for normal
galaxies. The accretion is highest in the beginning and dwindles as
the time goes by. Our result does not support theories of \mbh--\sig\
relation in which the black hole mass is a constant fraction of the
bulge mass/ velocity dispersion {\it at all times} or those in which
bulge growth is controlled by AGN feedback.

\end{abstract}


\section{Introduction}

Active galaxies are ``active'' because they accrete matter on to the
supermassive black holes. However, whether the accretion leads to
significant growth of the nuclear black hole has been a matter of some
debate. New results on X-ray background and the better determination
of local black hole mass density have led to the conclusion that
indeed, black holes grow during their active phase (e.g. \citet{bar01,
aller02, yu02}).

The above result needs to be understood in the context of the observed
tight correlation between the central black hole \mbh~ and the stellar
velocity dispersion $\sigma_*$ of the bulge in a galaxy
(e.g. \citet{geb00a, ferr00, merr01}). Measuring black hole masses
using stellar and gas dynamics (\citet{kor95} and references
therein), these authors found that log \mbh~= $a+b\times$log
($\sigma_*/\sigma_0)$ with \mbh~ in units of \msun~ and $\sigma_0$=200
\kms. The slopes $b$ of this relation vary between $b=3.75$
\citep{geb00a} and $b=5.27$ \citep{ferr00}.  Most recently,
\citet{tre02} estimated $b=4.02$ and $a=8.13$ for a sample of 31
nearby galaxies. Interestingly, the above relation for normal galaxies
also extends to active galaxies \citep{geb00b, ferr01}.

The above two results imply that the formation and growth of the
nuclear black hole and the bulge in a galaxy are intimately related,
and several theoretical models have attempted to explain the observed
\mbh -- $\sigma$ relation (e.g. \citet{haeh03, haeh98, adams01} and
\citet{king03}). To understand the link between the black hole and the
bulge, it is important to determine whether (a) black hole mass is a
constant fraction of the bulge mass, or bulge velocity dispersion, at
all times, or (b) during some accreting phase, the \mbh -- $\sigma$
relation is not followed by AGNs. The former can be obtained if, e.g.,
the growth of a black hole matches the growth of its surrounding bulge
exactly during a merger. Alternatively, feedback from a black hole can
limit the bulge growth. In case (b), the accretion history of a black
hole is not tied to the bulge growth.  It is still possible that
today's highly accreting AGNs do lie on the \mbh -- $\sigma$ relation
for normal galaxies, and may eventually leave the relation as their
BHs grow. Or, today's highly accreting AGNs may lie below the \mbh --
$\sigma$ relation for normal galaxies, and would eventually reach the
relation as the active BHs become dead. Indeed, in a recent model by
\citet{jordi03}, which explicitly couples accretion with the stellar
system in the bulge, the observed \mbh -- $\sigma$ relation is the final
relation at the end of the accretion process.


It is of interest, therefore, to follow the tracks of AGNs on the \mbh
-- $\sigma$ plane.  Since black holes would grow fastest with high
accretion rates, active galaxies with close to Eddington accretion are
perhaps the best candidates. At low redshift, a lot of observational
evidence suggests that narrow line Seyfert 1 galaxies (NLS1s; a
subclass of Seyfert galaxies with full width at half maximum of
H$\beta$ lines less than 2000 \kms\citep{ost85}) accrete at close to
Eddington rate (e.g. \citet{pound95, gru03b} and references there
in). \citet{mat01} argued that NLS1s do not follow the \mbh --
$\sigma$ relation. This, however, was only a suggestive result because
the sample size was small and neither the black hole masses nor the
velocity dispersions were accurately measured. \citet{wan02} confirmed
the Mathur \etal result with a sample of 55 AGNs (see also
\citet{bian03}).  On the other hand, \citet{wang01} have argued that
both BLS1s and NLS1s follow the \mbh -- $\sigma$ relation. Clearly, it
is important to test whether NLS1s occupy a distinct region in the \mbh
-- $\sigma$ plane using a large, homogenous sample. In this paper we
present our results based on a complete sample of 110 soft X-ray
selected AGN.

Note also that NLS1s are interesting objects as they occupy one
extreme end of the ``eigenvector 1'' relation of AGNs
(\citet{bor92}). The most widely accepted paradigm for NLS1s is that
they accrete at close to Eddington rate and have smaller black hole
masses for a given luminosity compared to BLS1s. Finding their locus
on the \mbh--\sig\ plane is therefore a worthwhile experiment anyway
as we will either find that occupy a distinct region compared to BLS1s
or that they don't. The first option is interesting for the reasons
discussed above. On the other hand if we find that NLS1s follow the
\mbh--\sig\ relation like the BLS1s, it has important implications
towards our understanding of the AGN phenomenon. As noted above, we
already have a good evidence for smaller BH masses of NLS1s, at a
fixed luminosity. If they follow the \mbh--\sig\ relation, it would
imply that NLS1s preferencially reside in galaxies with bulges of
smaller velocity dispersion. This would be a direct evidence for
dependence of AGN properties on their large scale galactic
environment.

In section 2 we discuss the sample selection and
the methodology to determine the black hole masses using the widths of
the H$\beta$ lines. Widths of the forbidden [OIII] lines are used as
surrogates for bulge velocity dispersion. In section 3 we present the
results and the discussion is in section 4.

\section{Estimates of \mbh\ and $\sigma$}

\subsection{The Sample}

 The sample contains all bright soft X-ray selected AGN from the ROSAT
All-Sky Survey (RASS, \citet{vog99}). The selection criteria are given
in \citet{tho98} and \citet{gru99}.  About half of these sources are NLS1s (51
objects) and 59 are broad line seyfert 1s (BLS1s).  NLS1s and BLS1s
show similar distribution in their redshifts, luminosities and
equivalent widths of H$\beta$ \citep{gru03a}. From this original
sample of 110 AGNs, we removed all the objects in which [OIII] lines
were severely unresolved leading to errors in the FWHM measurements,
with S$/$N$<3$. This left us with a sample of 75 AGNs, 32 NLS1s and 43
BLS1s.

\subsection{Black hole mass}
 
 Based on the reverberation mapping of the H$\beta$ line of 28 PG
 quasars, \citet{kas00} derived an empirical relation between the
 width of the H$\beta$ line and the central black hole mass.  They
 found that log \mbh~=5.17+log $R_{\rm BLR}$ + $2\times \rm (log
 FWHM(H\beta) - 3)$ with \mbh~ given in units of \msun~and
 FWHM(H$\beta$) given in units of \kms. $R_{\rm BLR}$ is the radius of
 the broad emission line region (BLR) and is larger for more luminous
 sources: log $R_{\rm BLR}$ = $1.52 + 0.70\times ({\rm log} \lambda
 L_{5100}-37)$ where $R_{\rm BLR}$ given in units of light days,
 L$_{5100}$ is the monochromatic luminosity at 5100\AA, and $\lambda
 L_{5100}$ is in units of Watts. This relation is well calibrated,
 albeit with some scatter, and can be used to estimate \mbh\ of active
 galaxies using FWHM(H$\beta$) and luminosity. Many authors have used
 it to derive \mbh\ of Seyfert galaxies and quasars (e.g. \citet{laor98}
 and \citet{wan99}) and we do the same (except for NGC 4051, which is known
 not to follow the radius luminosity relation, and so we use the BH
 mass as measured by reverberation \citep{pet00}). For our
 sample, FWHM(H$\beta$) and optical luminosities are given in
 \citet{gru03a}.

\subsection{Velocity dispersion}

The stellar velocity dispersion $\sigma_*$ in the bulge of a galaxy
can be measured by the widths of the CaII triplet absorption features
at 8498.0, 8542.1, and 8662.1 \AA. In a sample of 85 AGN,
\citet{nel95, nel96} found a moderately strong correlation between
$\sigma_*$ and FWHM([OIII]), the full width at half maximum of the
[OIII]$\lambda 5007$ line. Therefore, \citet{nel00} suggested that the
FWHM([OIII]) can be used as a surrogate of $\sigma_*$ with
$\sigma_*~\approx~\sigma_{\rm [OIII]}$ = FWHM([OIII])/2.35. This
result was confirmed by \citet{bor03}, who stated that the [OIII]
width can predict the black hole mass to a factor of 5 assuming the
\mbh -- $\sigma$ relation.  \citet{shi03} used the FWHM[(OIII)] as a
surrogate of the bulge stellar velocity dispersion $\sigma_*$ for a
large sample of AGN and concluded that it can be extended even to
redshifts as high as z$\approx$3. We also use FWHM([OIII]), given in
\citet{gru03a}, as an estimate of $\sigma_*$ for our sample.

\section{\label{results} Results}

\begin{figure}
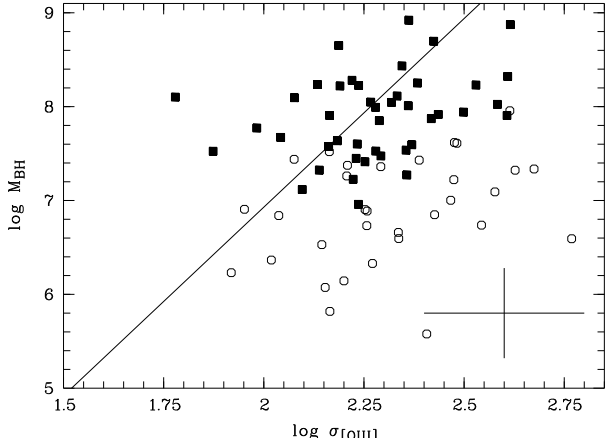

\clipfig{DGrupe.fig1}{85}{25}{14}{285}{192}
\caption{\label{sigma_mbh} Velocity dispersion $\sigma_{\rm [OIII]}$
vs. log $M_{\rm BH}$(H$\beta$). NLS1s are marked as open circles and
BLS1s as filled squares.  Black hole masses are given in units of
\msun.  The solid line marks the relation of \citet{tre02}. The cross
at the bottom right hand corner represents a typical error bar.  }
\end{figure}

Fig. \ref{sigma_mbh} plots the black hole mass derived from
FWHM(H$\beta$) vs. velocity dispersion \sig~derived from the
FWHM([OIII]) for our sample of 75 AGNs. The solid line is the relation
of \citet{tre02} obtained for normal galaxies. Clearly there is a
large scatter around this relation, and most likely implies that the
surrogates do not reproduce \mbh\ and \sig~ accurately. A comparison of
the black hole masses derived from our data with
reverberation mapping results (\citet{kas00, pet00,wan99,koll03})
shows that the black hole masses agree on average, with a random
scatter of about 0.2 dex.  Considering the scatter in the
radius--luminosity relation, variability and the unknown geometry of
the broad line region, we conservatively estimate the error on BH mass
measurement to be 0.5 dex.

 The observational error in $\sigma$ depends on the strength
of the [OIII] line. As shown in \citet{gru03b} there is a strong
anti-correlation between the FWHM([OIII]) and equivalent width of
[OIII]. Plus the weaker the [OIII] line the stronger the FeII emission
becomes (e.g. \citet{bor92, gru03b}) which makes the
line measurements in the objects with broader [OIII] more uncertain
than those with narrow [OIII] emission. The errors of the FWHM([OIII])
are given in \citet{gru03a} and are of the order of 0.2 dex. Clearly,
errors on both quantities, \mbh\ and \sig~ are large for individual
objects. However, we are interested in statistical differences in the
two populations, of a large number of BLS1s and NLS1s, and not on
exact values of individual sources.

Figure 1 shows that BLS1s and NLS1s occupy two distinct regions in the
\mbh -- \sig~ plane.  For a given velocity dispersion  NLS1s tend to show
smaller smaller black hole masses than BLS1s. If true, this is an important
result. However, before coming to that conclusion, we have to make
sure that the result is not spurious. A spurious result may be
obtained if the black hole masses of NLS1s are systematically
underestimated or if their velocity dispersions are systematically
overestimated relative to BLS1s. We will check these two cases below.

\begin{figure}
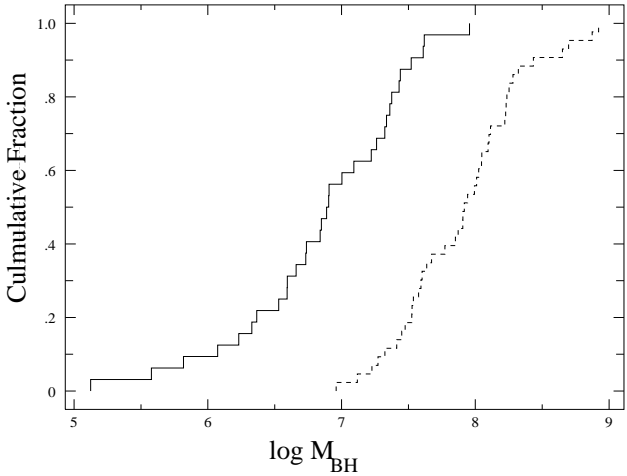

\clipfig{DGrupe.fig2}{85}{35}{75}{185}{195}
\caption{\label{distr_mbh} Cumulative fraction of a KS test of the 
black hole mass distributions of
NLS1s (solid line) and BLS1s (dashed line) given in units of \msun.
}
\end{figure}

{\it Are the black hole masses of NLS1s wrong?} Fig. \ref{distr_mbh}
shows the cumulative fraction of the distributions of the inferred
black hole masses of NLS1 (solid line) and BLS1s (dashed line) for a
Kolmogorov-Smirnov (KS) test.  The plot clearly shows that NLS1 and
BLS1s have different distributions of the black hole masses.  In
general, more luminous AGNs have higher black hole masses, but for a
given luminosity NLS1s have black hole masses about an order of
magnitude lower than BLS1s (note that the BLS1s and NLS1s in our
sample have similar luminosities; $\S 2.1$). This result confirms
earlier findings of, e.g. \citet{wan98} and \citet{pet00} and is
unlikely to be spurious. In fact, NLS1s have narrower broad emission
lines because of the smaller black hole masses. If they had masses
similar to the BLS1s, their BLRs would have to be relatively farther
away from the black hole. This, however, is not the case; \citet{pet00}
have found that NLS1s and BLS1s follow the same relation
between the BLR size and luminosity. We thus conclude that there is a
real difference in the black hole mass distribution of BLS1s and NLS1s.

\begin{figure}
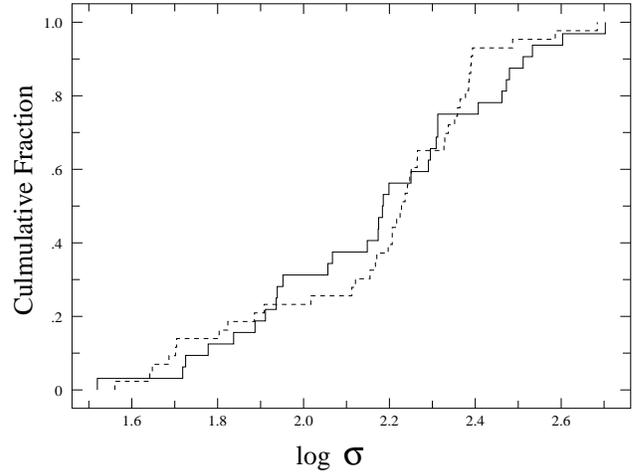

\clipfig{DGrupe.fig3}{85}{35}{76}{185}{195}
\caption{\label{distr_sigma} Cumulative fraction of a KS test of the
distributions of the stellar velocity dispersion $\sigma$ given in units of
\kms. The distribution of NLS1s is shown as a solid line and BLS1s are shown as
a dashed line.
}
\end{figure}

{\it Are the estimates of velocity dispersion wrong?} If FWHM([OIII])
is not a good surrogate, our estimates of \sig\ may well be wrong.
This could produce dichotomy between NLS1s and BLS1s if
$\sigma_{[OIII]}-\sigma_*$ is systematically different for the two
classes. The first indication that this is not the case comes from the
similarity of distribution of their \sig\ (Fig. \ref{distr_sigma}).  A
KS test shows that the two classes do not show any significant
difference. 

One might overestimate the [OIII] widths if the spectral resolution is
low and lines are not resolved. We correct for the instrumental line
broadening in our measurements \citep{gru03a}.
Moreover, if resolution was a problem, we would have seen a
clustering of line widths close to the instumental line
widths. Instead, we observe a wide range of widths, and the
distribution of widths for BLS1s and NLS1s is similar. This implies
that any problem with resolution is not artificially increasing the
line widths of NLS1s only. 

One problem which might affect NLS1s only is the strong FeII emission
close to O[III]$\lambda 5007$ emission line. For our entire sample,
FeII contribution has been subracted before making measurements on the
[OIII] lines \citep{gru03a}. If FeII contribution was
systematically undersubtracted, it will lead to overestimation of
[OIII] line widths. With this in mind, we re-examined the FeII 
subtracted spectra of all the NLS1s in our sample. We found that FeII
emission was not undersubtracted, and may even be slightly
oversubtracted, in all but but two cases. Deleting these two objects
with poorer S/N from our sample does not change the statistical result.

\begin{figure}
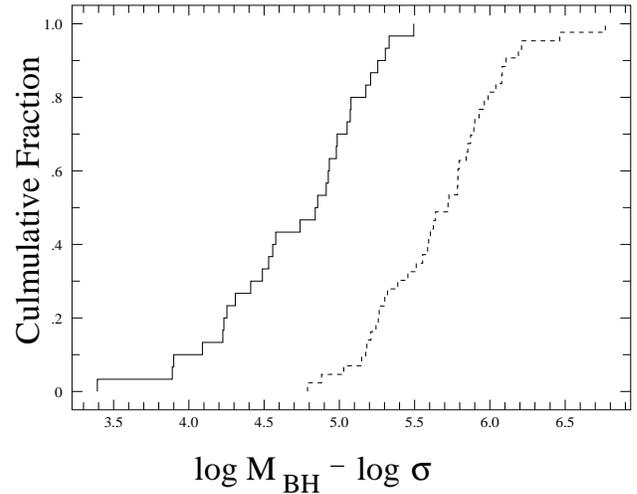

\clipfig{DGrupe.fig4}{85}{35}{73}{185}{192}
\caption{\label{distr_mbh_sigma} Cumulative fraction of a KS test of the
distributions of the black hole mass $M_{\rm BH}$ divided by the stellar 
velocity
dispersion $\sigma$.
The distribution of NLS1s is shown as a solid line and BLS1s are shown as
a dashed line.
}
\end{figure}

A more detailed look at many [OIII] lines shows that they have blue
asymmetry in a number of AGNs (e.g. \citet{gon99, lei00}),
possibly resulting from outflows. These observations need further
scrutiny since the asymmetry might have resulted in erronously large
measurements of FWHM([OIII]). (Note, however, that a part of the blue
asymmetry may be a result of oversubtracting the FeII emission from
the red part of the line.) To correct for this possible problem, we
re-measured the width of the [OIII] line as $2 \times$ half width at
half maximum of the red part of emission line.
While the new measurements reduce individual values of \sig, the BLS1s
 and NLS1s still ccupy distinct regions in the \mbh -- \sig~
 plot. This is clearly seen from figure\,\ref{distr_mbh_sigma} which
 plots the cumulative distribution of the ratio \mbh/\sig~ for BLS1s
 and NLS1s. The two classes are significantly different, with formal
 K-S test probability of being drawn from the same population $<
 0.001$.

\section{Discussion}

In this paper we present distribution of black hole masses and bulge
velocity dispersions for a comple sample of AGNs. Black hole masses
are estimated from a well calibrated relation between the width of the
H$\beta$ emission line, luminosity and \mbh. Bulge velocity
dispersion is estimated from the width of the narrow [OIII] emission
line. Neither of these are accuarate measurements of the two
quantities and errors on the values of \mbh\ and \sig\ for individual
objects are large. The results presented here are therefore
statistical, and compare the broad distributions of the two classes of
AGNs, the BLS1s and the NLS1s. 

We find that BLS1s and NLS1s occupy two distinct regions in the \mbh
-- \sig\ plane. This does not appear to be a result of systematically
underestimating black hole masses or systematically overestimating the
[OIII] line widths of NLS1s. This might be still a spurious result if
the narrow emission line region of NLS1s is somehow much closer to the
center of the galaxy and so does not trace the bulge velocity
dispersion. While we cannot rule out this possibility, there are no
observations supporting this fact either. 

After carefully investigating all the options, we come to the
conclusion that the result is unlikely to be spurious in that NLS1s
and BLS1s do have different ratios of black hole mass to bulge
velocity dispersion. So if BLS1s follow the \mbh -- \sig\ relatation
for normal galaxies, then NLS1s do not. Needless to say, this result
will have to be confirmed with accurate measurements of black hole
masses and bulge velocity dispertions. If confirmed, it has important
consequences towards our understanding of black hole formation and
growth. We find that the black hole mass is not a constant fraction of
the bulge mass or of bulge velocity dispersion {\it at all times}. In
other words, growth of a black hole by accretion does not match the
growth of its surrounding bulge exactly or the accretion process
itself does not increase the bulge velocity dispersion. Thus, our
results are also inconsistent with the models of \mbh--\sig\ relation
in which feedback from a black hole controls the growth of the
bulge. Our resuls support a scenario in which black holes grow by
accretion in well formed bulges, possibly after a major
merger. Perhaps, the surge in accretion is a result of that merger
itself. As they grow, they get closer to the \mbh--\sig\ relation for
normal galaxies. The accretion rate is highest in the beginning and
dwindles as the time goes by. This scenario is consistent with the
recent theoretical model of \citet{jordi03} in which the \mbh--\sig\
relation is the final state at the end of the accretion process. The
observation that the broad line Seyfert 1s lie close to the
\mbh--\sig\ relation then tells us that the black hole mass growth at
low accretion rate is not significant. This scenario is also
consistent with the proposal of \citet{gru99} and \citet{mat00} that
NLS1s are younger members of the Seyfert population.  Presumably, the
insight we have got from studying the Seyfert galaxies is also
applicable to quasars. To understand further the history of BH and
bulge growth, similar observations at high redshift would be valuable.


\acknowledgments We thank Jordi Miralda-Escud\'e, Brad Peterson, Rick
 Pogge, Greg Shields, David Weinberg and Bev Wills for helpful
 discussions.  This research has made use of the NASA/IPAC
 Extra-galactic Database (NED) which is operated by the Jet Propulsion
 Laboratory, Caltech, under contract with the National Aeronautics and
 Space Administration.

\clearpage


\end{document}

%% file: DGrupe_clipfig.tex
\def\clipfig#1{\def\lbracket{[}\def\testit{#1}%
    \ifx\testit\lbracket\let\next=\optclipfig\else\let\next=\stdclipfig\fi%
    \next{#1}}
%
\newcommand {\hclipfig} [7] {\clipfig[#7]{#1}{#2}{#3}{#4}{#5}{#6}}
%
\def\usemodepsfig {\global\def\cfmode{x}\typeout{*** set clipfig to PSFIG mode ***}}
\def\usemodeepsf  {\global\def\cfmode{}\typeout{*** set clipfig to EPSF mode ***}}
\def\useunitmm    {\global\def\cfunit{x}\typeout{*** set clipfig to use mm as unit ***}}
\def\useunitcm    {\global\def\cfunit{}\typeout{*** set clipfig to use cm as unit ***}}
\def\clipfigsettings {\ifx\cfmode\empty\def\ccfmode{EPSF }\else\def\ccfmode{PSFIG }\fi%
    \ifx\cfunit\empty\def\ccfunit{cm }\else\def\ccfunit{mm }\fi%
    \typeout{*** current clipfig settings: \ccfmode mode, using \ccfunit as unit ***}}
%
%
%
%
\def\stdclipfig#1#2#3#4#5#6{\ifx\cfmode\empty%
    \let\next=\eclipfig\else\let\next=\pclipfig\fi%
    \next{#1}{#2}{#3}{#4}{#5}{#6}}
\def\optclipfig#1#2]#3#4#5#6#7#8{\ifx\cfmode\empty%
    \let\next=\ehclipfig\else\let\next=\phclipfig\fi%
    \next{#3}{#4}{#5}{#6}{#7}{#8}{#2}}
%
%
%
\newcommand {\pclipfig}[6] {\ifx\cfunit\empty%
        \psfig{figure=#1.ps,width=#2cm,bbllx=#3cm,bblly=#4cm,bburx=#5cm,%
           bbury=#6cm,clip=}\else%
        \psfig{figure=#1.ps,width=#2mm,bbllx=#3mm,bblly=#4mm,bburx=#5mm,%
           bbury=#6mm,clip=}\fi}
\newcommand {\phclipfig}[7] {\ifx\cfunit\empty%
        \hspace{#7cm}\psfig{figure=#1.ps,width=#2cm,bbllx=#3cm,bblly=#4cm,%
           bburx=#5cm,bbury=#6cm,clip=}\else%
        \hspace{#7mm}\psfig{figure=#1.ps,width=#2mm,bbllx=#3mm,bblly=#4mm,%
           bburx=#5mm,bbury=#6mm,clip=}\fi}
%
%
%
\newcommand {\eclipfig}[6]{%
  \ifx\cfunit\empty\epsfxsize=#2cm\else\epsfxsize=#2mm\fi%
  \epsfclipon\epsfverbosetrue%
  \cfcmtopspts{#3}\cfllxi=\cftempi\cfllxf=\cftempf%
  \cfcmtopspts{#4}\cfllyi=\cftempi\cfllyf=\cftempf%
  \cfcmtopspts{#5}\cfurxi=\cftempi\cfurxf=\cftempf%
  \cfcmtopspts{#6}\cfuryi=\cftempi\cfuryf=\cftempf%
  \def\cfstra{\number\cfllxi.\number\cfllxf}%
  \def\cfstrb{\number\cfllyi.\number\cfllyf}%
  \def\cfstrc{\number\cfurxi.\number\cfurxf}%
  \def\cfstrd{\number\cfuryi.\number\cfuryf}%
  \hbox{\epsfbox[{\cfstra} {\cfstrb} {\cfstrc} {\cfstrd}]{#1.ps}}}
\newcommand {\ehclipfig}[7]{%
  \ifx\cfunit\empty\epsfxsize=#2cm\else\epsfxsize=#2mm\fi%
  \epsfclipon\epsfverbosetrue%
  \cfcmtopspts{#3}\cfllxi=\cftempi\cfllxf=\cftempf%
  \cfcmtopspts{#4}\cfllyi=\cftempi\cfllyf=\cftempf%
  \cfcmtopspts{#5}\cfurxi=\cftempi\cfurxf=\cftempf%
  \cfcmtopspts{#6}\cfuryi=\cftempi\cfuryf=\cftempf%
  \def\cfstra{\number\cfllxi.\number\cfllxf}%
  \def\cfstrb{\number\cfllyi.\number\cfllyf}%
  \def\cfstrc{\number\cfurxi.\number\cfurxf}%
  \def\cfstrd{\number\cfuryi.\number\cfuryf}%
  \ifx\cfunit\empty\hspace{#7cm}\else\hspace{#7mm}\fi%
  \hbox{\epsfbox[{\cfstra} {\cfstrb} {\cfstrc} {\cfstrd}]{#1.ps}}%
  \vspace{-1mm}}
%
%
%
\newdimen\cfllxi \newdimen\cfllyi  \newdimen\cfurxi  \newdimen\cfuryi
\newdimen\cfllxf \newdimen\cfllyf  \newdimen\cfurxf  \newdimen\cfuryf
\newdimen\cftemp \newdimen\cftempi \newdimen\cftempf
\newdimen\cfpspoint \cfpspoint=1bp
%
%
%
\newcommand{\cfcmtopspts}[1]{\ifx\cfunit\empty%
  \cftemp=#1cm\else\cftemp=#1mm\fi%
  \multiply\cftemp10\divide\cftemp\cfpspoint%
  \cftempf=\cftemp\divide\cftemp10\cftempi=\cftemp\multiply\cftemp10%
  \advance\cftempf-\cftemp}
%
%
\def\cfmode{}\def\cfunit{}\clipfigsettings
%